\documentstyle[psfig,pra,aps,twocolumn]{revtex}
\begin{document}
\author{
R. Derka$^{1,2}$, V. Bu\v{z}ek$^{2,3}$, and A.K. Ekert $^{1}$
}
\address{$^{1}$ Department of Physics, Oxford University, Parks Road, OX1 3PU
Oxford, U.K .\\
$^{2}$ Institute of Physics, Slovak Academy of Sciences,
D\'ubravsk\'a cesta 9, 842 28 Bratislava, Slovakia.\\
$^{3}$ Optics Section, The Blackett Laboratory,
Imperial College, London SW7 2BZ,  U.K.\\
}
\title{Universal Algorithm for Optimal Estimation of Quantum States from
Finite Ensembles}
\date{13 July 1997}
\maketitle
\begin{abstract}
We present a universal algorithm for the optimal quantum state
estimation of an arbitrary finite dimensional system. The algorithm
specifies a {\em physically realizable} positive operator valued
measurement (POVM) on a finite number of identically prepared systems.
We illustrate the general formalism by applying it to different
scenarios of the state estimation of $N$ independent and identically
prepared two-level systems (qubits).
\end{abstract}

Suppose we have $N$ quantum objects, each prepared in an unknown {\it
pure} quantum state described by a density operator $\hat
\rho=|\psi\rangle \langle \psi|$. What kind of measurement provides us
with the best possible estimation of $\hat \rho$?

Clearly, if we have an unlimited supply of particles in state $\hat
\rho$, i.e. when $N$ approaches infinity, we can estimate $\hat \rho$
with an arbitrary precision. In practice, however, only finite and
usually small ensembles of identically prepared quantum systems are
available.  This leads to an important problem of the optimal state
estimation with fixed physical resources.

A particular version of this problem has recently been addressed by
Massar and Popescu~\cite{Mas&Pop} who have analyzed extraction of
information from finite ensembles of spin-1/2 particles. They have
shown that there exists a fundamental limit on the fidelity of
estimation and they have constructed a POVM consisting of an {\it
infinite} continuous set of operators which attains this limit. It has
been suggested that the optimal {\it finite} POVM may exist, however,
so far no explicit construction of such finite POVM has been
provided. We solve this problem by describing an algorithm which gives
the {\it optimal and finite} POVM. We also provide a concise
derivation of a general bound on the fidelity of a state estimation of
arbitrary finite-dimensional systems (the Massar--Popescu bound
follows as a particular case). Finite POVMs can be, at least in
principle, constructed and the optimal state estimation procedures may
be applied in many areas of physics ranging from the optimal phase
estimation to quantum computation.

In fact, in the following we are solving a more general problem of
optimal estimation of unitary operations performed on quantum objects
(the state estimation follows as a special case).  Let us assume that
state $\hat \rho$ is generated from a reference state
$\hat\rho_0=|\psi_0\rangle\langle \psi_0|$ by a unitary operation
$U({\bf x})$ which is an element of a particular unitary
representation of some group G. Different ${\bf x}$ denote different
points of the group (e.g., different angles of rotation in the case of
the $SU(2)$) and we assume that all values of ${\bf x}$ are equally
probable.

Our task is to design the most general POVM, mathematically described
as a set $\{\hat O_r\}_{r=1}^ R$ of positive Hermitian operators such
that $\sum_r \hat O_r = \hat{1}$ \cite{Peres,Helstrom}, which when
applied to the {\it combined} system of {\em all} $N$ copies provides
us with the best possible estimation of $\hat \rho$ (and therefore
also of $U({\bf x})$).  We quantify the quality of the state
estimation in terms of the {\it mean} fidelity
\begin{eqnarray}
\bar f & = & \sum_r \int_G \!\! d{\bf x}\ {\bf Tr}[\hat O_r\
\overbrace{U({\bf x})\hat\rho_0 U^\dagger({\bf x})\otimes \dots
\otimes  U({\bf x})\hat\rho_0
U^\dagger({\bf x})}^{\bf N\ times}] \nonumber \\
& \times &
{\bf Tr}[ U({\bf x}) \hat \rho_0  U^\dagger ({\bf x}) \ U_r 
\hat \rho_0 U_r^\dagger],
\label{1}
\end{eqnarray}
which corresponds to a particular choice of a cost function
\cite{Helstrom} used in a context of detection and estimation theory.
The mean fidelity (\ref{1}) can be understood as follows: In order to
assess how good a chosen measurement is we apply it many times {\it
simultaneously} on {\it all} $N$ particles each in state $U({\bf x})
\hat \rho_0 U^\dagger ({\bf x})$.  The parameter ${\bf x}$ varies randomly
and isotropically \cite{foot1} over all points of the group $G$ during
many runs of the measurement.

For each result $r$ of the measurement, i.e., for each operator $\hat
O_r$, we prescribe the state $|\psi_r\rangle=U_r|\psi_0 \rangle$ which
represents our guess (i.e., estimation) of the original state.  The
probability of the outcome $r$ is equal to $ {\bf Tr}[\hat O_r\ U({\bf
x})\hat\rho_0 U^\dagger({\bf x})\otimes \dots \otimes U({\bf
x})\hat\rho_0 U^\dagger({\bf x})]$ while the corresponding fidelity of
the state estimation is $|\langle \psi_r | \psi \rangle |^2= {\bf Tr}[
U({\bf x}) \hat \rho_0 U^\dagger ({\bf x}) \ U_r \hat \rho_0
U_r^\dagger]$.  This fidelity is then averaged over all possible
outcomes and over many independent runs of the measurement with
randomly and isotropically distributed parameters ${\bf x}$. We want
to find the generalized measurement which {\it maximizes} the mean
fidelity $\bar f$ given by Eq.(\ref{1}).

The combined system of $N$ identically prepared reference states
always remains within the {\em totally symmetric subspace} of
$H^k\otimes H^k\otimes \dots H^k$, where $H^k$ is $k$-dimensional
Hilbert space of the reference state in which the corresponding
unitary representation $U({\bf x})$ acts. Thus the dimensionality $d$
of the space in which we construct the POVM $\{\hat O_r\}$ is $d=(\
^{N+k-1}_{k-1})$.  In this case 
the first trace in Eq.(\ref{1}) can be
rewritten as
\begin{eqnarray}
\bar f & = &\sum_r \int_G {\bf Tr}[\hat O_r\  U^{N}({\bf x})\hat \Omega_0 U^
{N\dagger}
({\bf x})] \nonumber \\
& \times & {\bf Tr}[ U({\bf  x})\hat \rho_0 U^\dagger({\bf  x})\ 
U_r\hat \rho_0 U_r^\dagger ] d{\bf x},
\label{2}
\end{eqnarray}
where $ U^{N}({\bf x})$ is a new representation of the same group $G$;
it is equivalent to the $N$-fold symmetrized direct product
\cite{Barut&Roncka} of the original representation $U({\bf x})$. 
Here
$ U^{N}({\bf x})$ transforms the $(\ ^{N+k-1}_{k-1})$-dimensional
reference state denoted as $\hat{\Omega}_0$.

We can insert the identity operator $U_{r}^N U^{N\dagger}_{r}$ into
the first trace in Eq.~(\ref{2}) and, taking into account that in
Eq.(\ref{2}) we integrate over whole the group $G$ parameterized by
${\bf x}$, we can substitute $U^N({\bf x})U^{N\dagger}_{r}\rightarrow
U^N({\bf x}) $ and $U({\bf x})U^\dagger_{r}\rightarrow U({\bf x})$.
Now, using the linearity of the trace operation as well as the
linearity of the representation of the group $G$ ($U\hat \rho U^\dagger$ is
a linear adjoint representation) we rewrite Eq.(\ref{2}) as
\begin{eqnarray}
\bar f=\sum_r\ {\bf Tr}[\hat O_r\  U_{r}^N \hat F U^{N\dagger}_r],
\label{3}
\end{eqnarray}
where
\begin{equation}
\hat F= \int_G \
 U^{N}({\bf x})\hat \Omega_0 U^{N\dagger} ({\bf x})\ \ {\bf Tr}[
U({\bf x})\hat \rho_0 U^\dagger ({\bf x})\ \hat \rho_0] d{\bf x},
\label{4}
\end{equation}
is a positive Hermitian operator. 

Let us now derive an upper bound on the mean fidelity.  Taking into
account positivity of the operator $\hat F$ (i.e., $\hat
F=\sum_i\lambda_i |\phi_i\rangle\langle \phi_i|;~~\lambda_i\geq 0$)
and the completeness condition for POVM (i.e., $\sum_r \hat O_r=\hat
1$) we obtain

\begin{eqnarray}
\bar f&\!\! = \!\!\! &\sum_r\ \!\!\! {\bf Tr}[\hat O_r U_{r}^N \hat F 
U^{N\dagger}_r]
	  \! = \!\! \sum_{i\ r}\!\!\!
	    \ \lambda_i {\bf Tr}[\hat O_r
	    U_{r}^N |\phi_i\rangle\langle \phi_i| U^{N\dagger}_r]
\nonumber \\
     &\leq & \lambda_{max}\sum_{i\ r}
	     \	{\bf Tr}[\hat O_r\
	     U_{r}^N |\phi_i\rangle\langle \phi_i| U^{N\dagger}_r]  
\label{5} \\
	   &\! = \! & \lambda_{max} \sum_r \!\!\!
	     \	{\bf Tr}[\hat O_r\  U_{r}^N \hat 1 U^{N\dagger}_r]
	      \! = \! \lambda_{max} \ {\bf Tr}[ \hat 1]
             =\lambda_{max}\ d.\nonumber
\end{eqnarray}
From Eq.(\ref{5}) it clearly follows that the upper bound can be
achieved if and only if all operators $\hat O_r$ forming the POVM
satisfy the following conditions:\newline {\it i)} Each $\hat O_r$ is
proportional to a suitably rotated (by some $U_{r}^N$) projector on
the eigenvector of $\hat F$ with the highest eigenvalue, i.e.  for all
$\hat O_r$ there exists $ U_{r}^N$, such that $ \hat O_r=c_r^2\
U_{r}^N|\phi_{max}\rangle\langle \phi_{max}|U^{N\dagger}_r$. This
$U_{r}^N$, or more precisely $U_r |\psi_0\rangle$, is our guess
associated with the result ``$r$''.\newline {\it ii)} All $c_r^2$ are
real and positive, to assure that all $\hat O_r$ are positive
operators.\newline {\it iii)} Finally, the operators $\hat O_r$ have
to satisfy the completeness criterion $\sum_r\ c_r^2\
U_{r}^N|\phi_{max}\rangle\langle \phi_{max}|U^{N\dagger}_r=
\hat 1$.\newline

However, these three conditions are not necessarily compatible (see
Example~B below).  Therefore, in general, the solution cannot be found
via this type of considerations and we have to proceed in a different
way.

We start with the following observation.  Let us assume that we have
some POVM $\{\hat O_r \}_{r=1}^R$ and the corresponding guesses
$U_{r}^N$ which maximize the mean fidelity $\bar f$. We can always
construct another POVM with more elements which is also optimal. For
example, let us consider a one-parametric subgroup $U(\phi)=\exp
(i\hat X \phi)$ of our original group $G$ and choose a basis $\{
|m\rangle \}_{m=1}^d$ in which the action of this subgroup is
equivalent to multiplication by a factor ${\bf e}^{i \omega_m
\phi}$ (i.e., the operator $U(\phi)$ is diagonal in this basis and
$\omega_m$ are eigenvalues of the generator $\hat X$).  Then we take
$d$ points $\phi_s$ ($s=1,\dots d$) and generate from each
original operator $\hat O_r$ a set of $d$ operators $\hat O_{rs}^,= {1
\over d} U^{N}(\phi_s) \hat O_r U^{N\dagger} (\phi_s)$. In this way we
obtain a new set of $(d\cdot R)$ operators such that the mean fidelity
for this new set of operators, $\bar f=\sum_{r,s}{\bf Tr}[\hat
O_{r,s}^,\ U_{r,s}^N \hat F U_{r,s}^{N\dagger}]$, is {\it equal} to
the mean fidelity of the original POVM $\{\hat O_r\}$ because we
ascribe to each eventual result $[r,s]$ a new guess $U_{r,s}=U(\phi_s)
U_r$.  However, in order to guarantee that the new set of operators
$\hat O_{rs}^,$ is indeed a POVM we have to satisfy the completeness
condition
\begin{eqnarray}
\hat 1 & = &\sum_{s} \sum_r \hat O^,_{rs}=
\sum_{s} \sum_r  {1 \over d} U^N(\phi_s)\hat O_r
U^{N\dagger} (\phi_s)  \nonumber \\
       & = &\sum_s \sum _{m,n} 
{{\bf e}^{i \phi_s(\omega_m-\omega_n)}\over d} \sum_r \left(
\hat O_r \right)_{mn}
|m\rangle \langle n|.
\label{8}
\end{eqnarray}
Let us notice that, 
by the appropriate choice of $\phi_s$, the sum
$\sum_s {{\bf e}^{i \phi_s(\omega_m-\omega_n)}\over d}$ can {\em
always} 
 be made equal to $\delta_{m,n}$ providing all eigenvalues
are non-degenerate \cite{foot-sm} 
(this is basically a discrete
Fourier transform and we illustrate this point in detail in
Example~A).
In this case, the conditions (\ref{8}) for the off-diagonal terms in
the basis $| m\rangle$ are {\em trivially} satisfied whereas the
diagonal terms are equal to unity because the original POVM $\{\hat
O_r\}$ guarantees that $\sum_r (\hat O_r )_{mm}=1$. Moreover, even if
the original set of operators $\{\hat O_r\}$ does not satisfy the full
completeness condition and the conditions for the off-diagonal terms
are {\it not} satisfied (i.e., these operators do not constitute a
POVM) we can, using our extension ansatz, always construct a {\it
proper } POVM $\{\hat O_{r,s}\}$.  This {\it proves} that when we
maximize the mean fidelity (\ref{3}) it is enough to assume $d$
diagonal conditions rather than the original complete set of $d^2$
constraints for diagonal {\it and} off-diagonal elements.

Now we turn back to our original problem of how to construct the POVM
which maximizes the mean fidelity.  To do so we first express the
operators $\hat O_r$ in the form $\hat O_r\!= \!c_r^2\ U_{r}^N
|\Psi_r\rangle \langle \Psi_r| U^{N\dagger}_r$, where $|\Psi_r
\rangle$ are general normalized states in the $d$-dimensional space in
which the operators $\hat O_r$ act, and $c_r^2$ are positive
constants.  This substitution is done without any loss of generality
\cite{foot2} and it permits us to rewrite Eq.(\ref{3}) so that the
mean fidelity $\bar f$ does not explicitly depend on $U_r^N$, i.e.
\begin{equation}
\bar f=\sum_r\ c_r^2\ {\bf Tr}[|\Psi_r\rangle \langle \Psi_r| \hat F ].
\label{6}
\end{equation}
Obviously, the completeness condition $\sum_r \hat O_r=\hat 1$ is now
modified and it reads
\begin{equation}
\sum_r	c_r^2\ U_{r}^N |\Psi_r\rangle \langle \Psi_r|  
U^{N\dagger}_r\  =\  \hat 1.
\label{7}
\end{equation}  

From our discussion above it follows that when maximizing the mean
fidelity (\ref{6}) it is enough to apply only $d$ constraints $\sum_r
c_r^2\left|\langle m |U_{N,r}|\Psi _r\rangle \right|^2 =1$ (here
$m=1,\dots, d$) out of the $d^2$ constraints (\ref{7}). Therefore to
accomplish our task we solve a set Langrange equations with $d$
Lagrange multipliers $L_m$.  If we express $L_m$ as eigenvalues of the
operator $\hat L =\sum_m L_m\ |m\rangle \langle m|$ then we obtain the
final very compact set of equations determining the optimal POVM
\begin{eqnarray}
\left[ \hat F - U^{N\dagger}_r\! \hat L U_r^N \right]\! |\Psi_r\rangle\! 
=\! 0, &\ \ \ \ &
\sum_r \! c_r^2\left|\langle m |U_r^N|\Psi _r\rangle \right|^2\!\!\! =\! 1.
\label{9}
\end{eqnarray}
From here it follows that $|\Psi_r\rangle$ are determined as
zero-eigenvalue eigenstates.  More specifically, they are functions of
$d$ Lagrange multipliers $\{L_m\}_{m=1}^d$ and $R$ vectors $\{{\bf
x}_r\}_{r=1}^R$ [where ${\bf x}_r$ determine $U_r$ as $U_r=U({\bf
x}_r)$].  These free parameters are in turn related via $R$ conditions
${\bf Det}[ (\hat  F - U_r^{N\dagger} \hat L U_r^N)]=0 $.  The mean
fidelity now is equal to ${\rm Tr} \hat L$.  At this stage we solve a
system of $d$ linear equations [see the second formula in
Eq.(\ref{9})] for $R$ unknown parameters $c_r^2$.  All solutions for
$c_r^2$ parametrically depend on $L_m$ and ${\bf x}_r$ which are
specified above.  We note that the number of free parameters in our
problem depends on $R$ which has not been specified yet. We choose $R$
such that there are enough free parameters so that the mean fidelity
is maximized and simultaneously all $c_r^2$ are positive.  This
freedom in the choice of the value of $R$ also reflects the fact that
there is an infinite number of equivalent (i.e., with the same value
of the mean fidelity) optimal POVMs.  The whole algorithm is completed
by finding $\phi_s$ from Eq.(\ref{8}) which explicitly determine the
{\it finite} optimal POVM $\{ \hat O_{rs}^,\}$. This is the main
result of our Letter.

In the following we will apply this general algorithm into two simple
examples:\newline
{\it Example A}\newline 
Suppose we have $N$ 
identical copies of spin
$1/2$ all prepared in the same but unknown pure quantum state. If we
chose the group $G$ to be $U(2)$, i.e.  the complete unitary group
transforming a two-level quantum system, we can straightforwardly
apply the optimal estimation scheme as described above.  To be more
precise, due to the fact that there exist elements of the group $U(2)$
for which the reference state is the fixed point (i.e., it is
insensitive to its action) we have to work only with the coset space
$\ ^{SU(n)}|_{U(n-1)}$~\cite{Barut&Roncka}.  In the present case this
is a subset of the $SU(2)$ group parameterized by two Euler angles
$\theta,\psi$ (the third Euler angle $\chi$ is fixed and equal to
zero). This subset  is isomorphic to the Poincare sphere.

The unitary representation $U$ is now the representation $({1\over
2})$ (we use a standard classification of $SU(2)$ representations,
where $(j)$ is the spin number).  Its $N$-fold symmetrized direct
product (we denote this representation as $U^N$) is the representation
classified as $({N \over 2})$ (which transforms a spin-$N/2$
particle).  Choosing the standard basis $|j,m\rangle$ with $m=-j,\dots
j$ in which the coordinate expression for $U(\theta,\psi)$ corresponds
to standard rotation matrices $D^j_{m,n}(\theta,\psi,0)={\rm
e}^{-im\psi}\ d^j_{m,n}(\theta)$~\cite{Zare}, we obtain the matrix
expression for the operator $\hat F$
\begin{eqnarray}
F_{m,n} & = &  \int _0^{2\pi} d\phi \int_0^\pi {\sin (\theta)
d\theta\over 8\pi}  (1 +\cos \theta)
\label{10} \\
 & \times &
D^{N \over 2}_{m,{N \over 2}}(\theta,\phi)
\ D^{{N \over 2}\ast }_{n,{N \over 2}}(\theta,\phi)
 =  {N/2 + m + 1 \over (N + 2)(N + 1)}\delta_{m,n}.
\nonumber
\end{eqnarray}
When we insert this operator in the Eq.(\ref{5}) we immediately find
the upper bound on the mean fidelity to be equal to ${N+1\over N+2}$.
This is the result derived by Massar and Popescu \cite{Mas&Pop} who
have also shown that this upper bound can be attained using the
special POVM which consists of an {\it infinite} continuous set of
operators proportional to isotropically rotated projector $|{N \over
2},{N \over 2} \rangle\langle {N \over 2},{N \over 2}|$ \cite{foot3}.

However, following our algorithm, we can now construct the optimal
POVM which is {\it finite}. To do so, we have to find a finite set of
pairs of angles $\{ (\theta_r, \psi_r) \}$ such that the completeness
conditions (\ref{7}) which now take the form
\begin{equation}
\sum_{r} c^2_r\  {\rm e}^{-i\psi_r(m-n)} d_{m,{N \over 2}}^{N \over 2}
(\theta_r)\
 d_{n,{N \over 2}}^{N \over 2}(\theta_r) = \delta_{m,n}, 
\label{11}
\end{equation}
are fulfilled.  Following our general scheme we first satisfy the
completeness conditions (\ref{11}) for diagonal terms [compare with
Eq.(\ref{9})]
\begin{eqnarray}
\sum_r	c^2_r\	d_{m,{N \over 2}}^{N \over 2}(\theta_r)^2= 1;\qquad
 m=-N/2,\dots	N/2.
\label{12}
\end{eqnarray}
To satisfy these completeness conditions we choose $N+1$ angles
$\theta_r$ to be equidistantly distributed in the $\langle 0,\pi
\rangle$ (obviously, there are many other choices which may suite the
purpose -- see discussion below Eq.(\ref{9})).  Then we solve the
system of linear equations for $N+1$ variables $c_r^2$.  For this
choice of $\theta_r$ the system~(\ref{12}) has non-negative solutions.
Finally we satisfy the off-diagonal conditions by choosing $N+1$
angles $\psi_{s}={2 s\pi \over N+1}$ for each $\theta_r$. In this case
${1 \over N+1} \sum_{s=0}^N {\rm e}^{i \psi_s y}=\delta_{y,0}$ for all
$y=-N/2,\dots N/2$ and the off-diagonal conditions are satisfied
straightforwardly.  This concludes the construction of the {\em
optimal} and {\it finite} POVM for the spin-$1/2$ state estimation.

{\it Example B}\newline The algorithm can also be used to estimate a
unitary evolution of quantum systems.  Consider, for example, a system
of $N$ qubits, all initially prepared in some reference state and
undergoing a free evolution described by the $U(1)$ group (e.g. like
in the Ramsey type experiments~\cite{Ramsey}).  Our task is to find a
measurement which provides the optimal estimation of the phase of the
$U(1)$ rotation. N.B. the phase estimation without any a priori
information is different from the frequency standards experiments,
where the issue is the ability to distiguish neighbouring quantum
states with the best resolution (see, for example~\cite{Wineland}).

All unitary irreducible representations of $U(1)$ are one dimensional
and they are parameterized by a single integer number $(j)$.  Rotation
of a single isolated qubit is described by a representation $U$
classified as $(0)\oplus(1)$ (here $\oplus$ denotes a direct sum of
representations). The entire system of $N$ qubits (these are assumed
to be nonentangled) is then transformed due to the representation
$U^N$ which is equal to the symmetrized $N$-fold direct product of the
basic representation $U$.  The representation $U^N$ is equal to a
direct sum of irreducible representations of the form $(0)\oplus
(1)\oplus \dots (N)$ which act in the $N+1$ dimensional space spanned
by basis vectors $|m\rangle$, $m=0,1,\dots N$.  In this basis matrix
elements $\hat F_{m,n}$ of the operator $\hat{F}$ given by Eq.(\ref{4})
take the form
\begin{eqnarray}
\hat F_{m,n}  &=& \int_0^{2\pi}{d\psi \over 2\pi}\ {\sqrt{
(^N_{N-m})
(^N_{N-n})}\over 2^{N+1}}\
{\rm e}^{i\psi (n-m)}\ (1 + \cos\psi) \nonumber \\
& =& {\sqrt { (^N_{N-m})
(^N_{N-n})} \over 2^{N+2}} \left( 2 \delta_{m,n} +
 \delta_{m,n+1} +  \delta_{m+1,n} \right).
\end{eqnarray}
The upper bound on the fidelity Eq.~(\ref{5}) is now too conservative
to be of any use (greater than unity).  We can, however, solve the
system of Eqs.~(\ref{9}) which in this particular case reads
\begin{eqnarray}
\left[ \hat F -\hat L \right ]|\Psi \rangle=0, \qquad
|\langle m|\Psi\rangle|^2=1; 
\ \ \ \forall m.
\label{14}
\end{eqnarray}
The condition ${\bf Det}(\hat F - \hat L)=0$ now determines the
eigenvector $|\Psi\rangle$ with the zero eigenvalue as a function of
Lagrange multipliers $L_m$.  When we substitute this eigenvector into
the second equation in Eq.(\ref{14}) we obtain a set of equations for
$L_m$ from which the reference state $|\Psi \rangle$ can be
determined.  The final POVM is then constructed by rotation of
$|\Psi\rangle$ by $N+1$ angles $\phi_s$ in such a way that all
off-diagonal elements of $\sum_s(\hat O_s)_{m,n}$ 
become equal to zero. This is done
in exactly the same way as in Example~A.  The resulting POVM
corresponds to the {\it von Neumann measurement} performed on the {\em
composite} system of {\em all} $N$ ions.

We note that the mean fidelity $\bar f$	has a rapidly growing number
of local extrema which originate from the roots of certain
polynomials. From this set of extrema we easily choose the global
maximum which corresponds to the optimal POVM.

The maximal mean fidelity for the first six $N$ is: ${3
\over 4}$, ${{2+\sqrt 2} \over 4}$, ${{11 + 2 \sqrt 3 }\over 16}$,
${{5 + \sqrt 6} \over 8}$, ${{32+ 5 \sqrt 2 + 2 \sqrt 5 + 5 \sqrt 2
\sqrt {3 + 2 \sqrt 2}} \over 64}$, ${{32+ 10 \sqrt 3 + 3 \sqrt {10} +
\sqrt 6} \over 64}$,$\dots$ which approximately gives: $0.750$,
$0.854$, $0.904$, $0.931$, $0.947$, $0.957,\dots$

In conclusion, we have presented a general algorithm for the optimal
state estimation from finite ensembles. It provides finite POVMs
which, following the Neumark theorem~\cite{Neumark}, can, at least in
principle, be implemented as simple quantum computations.

We thank Serge Massar, Susana Huelga, Thomas Pellizzari and Chiara
Macchiavello for helpful discussions.  This work was supported by the
Open Society Fund and FCO, the United Kingdom EPSRC, European TMR
Network ERP-4061PL95-1412, Hewlett-Packard, Elsag-Bailey, The Royal
Society, and the Grant Agency VEGA of the Slovak Academy of Sciences.


\begin{references}
\bibitem{foot1}
We note that this isotropy condition is equivalent to a ``no a priori
information'' condition and is associated with the specific
integration measure in Eq.(\ref{1}). This measure has to be invariant
under the action of all unitary transformations on the state space of
pure states.
\bibitem{foot-sm}
In the case when the spectrum of the generator $\hat X$ is degenerate,
i.e., for some $m$ and $n$ we have $\omega_m=\omega_n$, then our algorithm
is still valid, provided we increase a number of Langrange multipliers
in Eq.(\ref{9}) to account for off-diagonal elements $L_{mn}$ and $L_{nm}$
in the definition of the operator $\hat L$ in Eq.(\ref{9}).
\bibitem{foot2}
The most general choice of $\hat O_r$ would be
$\hat O_r=\sum_i c^2_{r,i}\ |\Psi_{r,i}\rangle \langle
\Psi_{r,i}|$.
 However, from the point of view of optimality 
of the POVM these operators
are always less effective than operators 
$\hat O_r\!= \!c_r^2\ U_{r}^N |\Psi_r\rangle \langle \Psi_r| U^{N\dagger}_r$
which are proportional to one-dimensional projectors.
\end{references}
\end{document}